\documentclass[prd,twocolumn,showpacs,amsmath,amssymb,preprintnumbers]{revtex4}
\usepackage{amssymb}
\usepackage{mathrsfs}
\usepackage{txfonts}
\usepackage{hyperref}
\usepackage{graphicx}
\usepackage{dcolumn}
\usepackage{bm}
\usepackage{xcolor}
\usepackage{soul}
\usepackage{amsmath}
\begin{document}

\title{Dynamics and observational signatures of warm Dirac-Born-Infeld inflation with nonminimal derivative coupling}

\author{Run-Qing Zhao}
\email{18733666528@163.com}
\affiliation{School of Science, Qingdao University of Technology, Qingdao 266033, China}
\author{Xiao-Min Zhang}
\thanks{Corresponding author}
\email{zhangxm@mail.bnu.edu.cn}
\affiliation{School of Science, Qingdao University of Technology, Qingdao 266033, China}
\author{Peng-Cheng Chu}
\email{kyois@126.com}
\affiliation{School of Science, Qingdao University of Technology, Qingdao 266033, China}
\author{Yun-Cai Feng}
\email{fengyuncai@qut.edu.cn}
\affiliation{School of Science, Qingdao University of Technology, Qingdao 266033, China}
\date{\today}

\begin{abstract}

This paper investigates a warm Dirac-Born-Infeld (DBI) inflationary model with nonminimal derivative coupling (NMDC) to gravity, where the inflaton kinetic term interacts with the Einstein tensor, thereby improving the effective gravitational friction. This model seamlessly integrates the noncanonical DBI kinetic structure, the NMDC-induced gravitational friction, and thermal dissipation. We formulate the background evolution equations along with the corresponding slow-roll stability conditions, leading to analytic results for the scalar spectral index $n_s$ and the tensor-to-scalar ratio $R$.
By applying these results to power-law potentials with $n=2$ and $n=4$, the model parameter space is constrained using the \emph{Planck} 2018 data. The findings indicate that the interaction between NMDC-induced gravitational friction and thermal dissipation effectively modulates $n_s$ and significantly expands the viable parameter space. In the $(n_s,R)$ plane, the predictions for $N=50$ fall within the 95\% confidence-level region, while those for $N=60$ extend into the 68\% confidence-level region and approach the observationally preferred central values. For the representative parameter choices examined, the tensor-to-scalar ratio is notably suppressed, generally within the range $10^{-8}\lesssim R\lesssim10^{-5}$.
Moreover, the combined damping mechanism relaxes the slow-roll condition related to $\eta$ and limits the inflaton field excursion, thus addressing the $\eta$ problem without incurring super-Planckian field variations. These results indicate that warm DBI inflation with NMDC offers a theoretically coherent and observationally viable inflationary model, showcasing the complementary effects of thermal dissipation and enhanced gravitational friction in the context of modified gravity.

\end{abstract}

\maketitle

\section{\label{sec:level1}Introduction}
The inflationary model is founded on a quasi-exponential accelerated expansion in the very early Universe, effectively addressing multiple issues present in the standard cosmological framework, such as the horizon, flatness, and monopole problems \cite{Guth1981, Linde1982, Albrecht1982, Bassett2006}. Meanwhile, this theory provides a natural physical explanation for the formation of the Universe's large-scale structure and the minute anisotropies observed in the cosmic microwave background (CMB) through vacuum fluctuations \cite{PLANCK1, PLANCK2}.

Inflationary theory can be classified into two main paradigms: standard (cold) inflation and warm inflation. Warm inflation, first proposed by A. Berera in 1995 \cite{BereraFang, Berera1995}, retains the success of cold inflation in resolving the horizon, flatness, and monopole problems, while exhibiting distinct physical properties. Most notably, thermal dissipation considerably relaxes the slow-roll conditions \cite{Ian2008, Campo2010, ZhangZhu2013, Zhang2014}, which effectively addresses the ``$\eta$ problem'' \cite{etaproblem, etaproblem1} faced by cold inflation models and mitigates the concern over excessively large inflaton field amplitudes \cite{Berera2006, BereraIanRamos2009}.
The crucial difference between cold and warm inflation lies in the source of cosmological density fluctuations: in warm inflation, these fluctuations stem predominantly from thermal fluctuations \cite{BereraIanRamos2009, Berera2000, Lisa2004, Taylor2000, Chris2009}, whereas in cold inflation, they originate from vacuum quantum fluctuations \cite{Bassett2006, LiddleLyth}. During warm inflation, the inflaton interacts with subdominant bosonic or fermionic fields, producing radiation continuously. This interaction introduces a thermal damping term that slows the inflaton's evolution, allowing the Universe to transition smoothly into the radiation-dominated Big Bang epoch without requiring a separate reheating phase after inflation.

Inflation is conventionally modeled using a canonical scalar field, with a Lagrangian density given by $\mathcal{L}=X-V$, where $X=\frac12 g^{\mu\nu}\partial_{\mu}\phi\partial_{\nu}\phi$ and $V$ represents the inflaton potential. Beyond this fundamental theoretical framework, numerous extended formulations have been developed to establish inflationary models, including noncanonical field models, purely kinetic field models, multifield models \cite{refining2012,Mukhanov2006,Armendariz-Picon1999,Garriga1999,Gwyn2013,Tzirakis,Franche2010,Easson2013,Bean2008,multifield}, and nonminimal coupling field models, among others \cite{Kaiser1995,Karydas2102.08450,GermaniPRL2010}.
Over the past three decades, warm inflation theory has experienced considerable advancement, with significant progress made in various research areas, including its microphysical foundations \cite{Berera2016,MossXiong2006}, cosmological perturbations \cite{Berera2000,Chris2009,Berera2016,WangYY2019,Zhang2023,WarmSPy2024,Moss2007}, and extended model frameworks \cite{Zhang2014,Zhang2018,Peng2016,WangYY2018,EadkhongNPB2023}.

Though inflationary models within the general relativity (GR) framework have been thoroughly investigated, research on inflation in modified gravity frameworks has predominantly concentrated on cold inflation models \cite{Kaiser1995,Karydas2102.08450,GermaniPRL2010,EadkhongNPB2023,DalianisJCAP2020}, with relatively limited exploration of warm inflation scenarios. The initial effort to extend warm inflation to nonminimal gravitational coupling theory was introduced in \cite{Sadjadi2015}, and the nonminimal derivative coupling (NMDC) noncanonical warm inflation was elaborated in the subsequent Ref. \cite{Zhang2024}.

In contrast to conventional nonminimal coupling models that directly couple the inflaton field, this framework establishes a nonminimal coupling between the inflaton and the gravitational sector through the inflaton's kinetic term \cite{Karydas2102.08450,GermaniPRL2010,DalianisJCAP2020,Sadjadi2015}. The NMDC term $G^{\mu\nu}\partial_{\mu}\phi\partial_{\nu}\phi$ introduced in the framework enhances the gravitational friction effect, which decelerates the evolution of the scalar field. By incorporating nonminimal coupling gravitational friction or thermal effects that yield the friction term $\Gamma\dot{\phi}$, the $\lambda\phi^4$ potential case, previously ruled out in minimal-coupling standard inflation, can be aligned with observational data \cite{YangNan2015,BeingWarm2014}.

In our earlier work \cite{Zhang2024}, we integrated the NMDC mechanism into noncanonical warm inflation, while Ref.~\cite{Cai2011} was the first to implement warm Dirac-Born-Infeld (DBI) inflation in general relativity. Nevertheless, a unified framework that combines DBI dynamics, NMDC-induced gravitational friction, and thermal dissipation has yet to be studied. This paper fills this gap by establishing a warm DBI inflation model in the NMDC framework, and systematically analyzes how the combined effects of these three ingredients modulate the background evolution, slow-roll dynamics, and observational predictions.
We systematically formulate the background dynamical evolution equations, perform linear stability analysis, and explore primordial perturbation theory. Furthermore, we analyze in detail how the combined influences of NMDC gravitational friction, the DBI noncanonical kinetic term, and thermal dissipation shape the inflationary background dynamics, slow-roll conditions, and primordial perturbation observables. Notably, the pronounced overdamping behavior facilitates the fulfillment of the slow-roll conditions and effectively addresses the $\eta$ problem. Utilizing the \textit{Planck} 2018 data \cite{PLANCK1}, we carry out extensive observational fitting and parameter constraints, confirming the model's strong compatibility with current cosmological observations. Our findings indicate that the competitive and synergistic interactions between NMDC gravitational friction and thermal dissipation not only significantly expand the viable parameter space of warm DBI inflation but also reduces the inflaton field excursion to well below the Planck scale ($\Delta\phi \ll M_p$), while substantially suppressing the tensor-to-scalar ratio, further improving the model's consistency with observational data.

The organization of this paper is as follows: Section \ref{sec:level2} reviews the foundational framework of canonical warm inflation and the essential theory of NMDC noncanonical warm inflation. Section \ref{sec:level3} develops a warm DBI inflation model in the NMDC framework and completes the derivation of background dynamics, linear stability analysis, and calculations pertaining to primordial perturbation theory. Section \ref{sec:level4} conducts observational fitting and parameter analysis of the model using \emph{Planck} 2018 data. Finally, Section \ref{sec:level5} summarizes the principal findings of this paper and suggests potential future research directions.

\section{\label{sec:level2}Warm inflation foundations and NMDC noncanonical warm inflation models}

\subsection{Original warm inflation}

Since the proposal of warm inflation, numerous models have been continuously proposed. Most of these models are based on canonical scalar fields, including those with a monomial potential \cite{BereraIanRamos2009}, a hybrid potential \cite{BereraIanRamos2009}, a hilltop potential \cite{BereraIanRamos2009,Sanchez2008}, and natural warm inflation \cite{Mishraa2012}, among others.

Using the conventional canonical scalar field as the inflaton, we outline the dynamics and fundamental equations of warm inflation. The total action for the matter sector of the universe is expressed as:
\begin{equation}
S = \int d^4x \, \sqrt{-g} \, \Big[ \mathcal{L}(X, \phi) + \mathcal{L}_R + \mathcal{L}_{\rm int} \Big] ,
\end{equation}
where $\mathcal{L}(X, \phi)$ signifies the Lagrangian density of the inflaton field, with $X = \frac{1}{2} \dot{\phi}^2$ defined in the spatially flat Friedmann-Robertson-Walker (FRW) universe. $\mathcal{L}_R$ represents the Lagrangian density of the radiation field, and $\mathcal{L}_{\rm int}$ denotes the interaction term between the inflaton and other sub-dominant fields.

In the FRW universe, for a homogeneous field $\phi$, the equation of motion for the canonical inflaton field can be written as:
\begin{equation}
\ddot{\phi} + (3H + \Gamma) \dot{\phi} + V_{\rm eff, \phi} = 0 ,
\label{motion_for_the_canonical}
\end{equation}
where $H = \frac{\dot{a}}{a}$ is the Hubble parameter, and $V_{\rm eff}$ denotes the effective potential with thermal corrections incorporated, with $V_{\rm eff,\phi}$ signifying its first-order derivative with respect to the inflaton field $\phi$. For simplicity, we refer to $V_{\rm eff}$ as $V$ hereinafter. Additionally, the parameter $\Gamma$ acts as the thermal dissipative coefficient for characterizing the thermal damping effect of the inflaton decaying into radiation, and it can be defined as either a constant or a function of the inflaton field and cosmic temperature \cite{Berera2016,MossXiong2006}.

A key parameter that measures the thermal dissipation strength in warm inflation is defined as:
\begin{equation}
r = \frac{\Gamma}{3H},
\end{equation}
where $r\gg 1$ and $r\ll 1$ indicate strong and weak dissipative warm inflation, respectively. The dissipative process of the inflaton field in warm inflation is associated with an increase in entropy density, which necessitates a thermodynamic description. The entropy density in warm inflation is expressed as
$s = -\frac{\partial f}{\partial T},$
where $f$ denotes the free energy density related to the energy density $\rho$ by
$f = \rho - Ts.$
Given that the warm inflation process is primarily influenced by the scalar field potential, the entropy density can be approximated as $s \simeq - V_T$, with $V_T$ denoting the derivative with respect to the temperature.

The total energy density and pressure of the universe in warm inflation are represented as:
\begin{equation}
\rho = \frac{1}{2} \dot{\phi}^2 + V(\phi,T) + T s,
\end{equation}
\begin{equation}
p = \frac{1}{2} \dot{\phi}^2 - V(\phi,T).
\end{equation}
Starting from the energy conservation equation of the FRW universe $\dot{\rho} + 3H(\rho+p) = 0$, combined with the field equation of motion given in Eq.~\eqref{motion_for_the_canonical}, the entropy production equation is derived as:
\begin{equation}
T\dot{s} + 3 H T s = \Gamma \dot{\phi}^2,
\label{entropy production}
\end{equation}
where the radiation energy density is defined as $\rho_r = 3T s/4$, and Eq.~\eqref{entropy production} is equivalent to $\dot{\rho}_r + 4 H \rho_r = \Gamma \dot{\phi}^2$.

The slow-roll approximation for warm inflation resembles that for cold inflation: it involves disregarding higher-order terms in the full equation of motion, and slow-roll inflation is universally potential-dominated. When applying the slow-roll approximation, the main equations of warm inflation are simplified to:
\begin{equation}
\dot{\phi} = - \frac{V_\phi}{3 H (1+r)},
\end{equation}
\begin{equation}
T s = r \dot{\phi}^2,
\label{eq:Ts}
\end{equation}
\begin{equation}
H^2 = \frac{8 \pi G}{3} V.
\end{equation}

To characterize and confirm the validity of the slow-roll approximation, a set of slow-roll parameters is defined for warm inflation. The three parameters describing the field dependence of $\phi$ are:
\begin{equation}
\epsilon = \frac{M_p^2}{2} \left( \frac{V_\phi}{V} \right)^2, \quad
\eta = M_p^2 \frac{V_{\phi\phi}}{V}, \quad
\beta = M_p^2 \frac{V_\phi \Gamma_\phi}{V \Gamma},
\label{slow-roll_parameters}
\end{equation}
where $M_p^2 \equiv (8\pi G)^{-1}$ denotes the squared reduced Planck mass. Two additional parameters related to temperature dependence are as follows:
\begin{equation}
b = \frac{T V_{\phi T}}{V_\phi}, \quad
c = \frac{T \Gamma_T}{\Gamma}.
\label{bc}
\end{equation}
Based on the stability analysis of warm inflation, the following slow-roll conditions can be obtained:
$\epsilon \ll 1 + r\ ;
\eta \ll 1 + r\ ;
\beta \ll 1 + r\ ;
b \ll \frac{r}{1+r}\ ;
|c| < 4$\ \cite{BereraIanRamos2009,Lisa2004,Ian2008,Campo2010}. This set of slow-roll conditions is considerably easier to satisfy compared to those in cold inflation, particularly in the strong dissipation regime with $r \gg 1$. To address the horizon and flatness problems, warm inflation also requires a sufficiently large number of e-folds $N \gtrsim 60$. The number of e-folds in warm inflation is expressed as:
\begin{equation}
N = \int H \, dt = - \frac{1}{M_p^2} \int_{\phi_*}^{\phi_{\rm end}} \frac{V}{V_\phi} (1 + r) \, d\phi',
\end{equation}
where $\phi_*$ indicates the field value of the inflaton at Hubble horizon crossing, and $\phi_{\rm end}$ represents the field value at the end of inflation.

\subsection{Noncanonical warm inflation models with nonminimal derivative coupling}

Most current research on warm inflation centers around canonical scalar fields. In contrast, noncanonical scalar fields, which represent a broader category, have been extensively investigated in cold inflation \cite{Franche2010,refining2012,Mukhanov2006,Armendariz-Picon1999,Garriga1999,Gwyn2013,Tzirakis,Easson2013,Bean2008}.
In our earlier work \cite{Zhang2024}, we expanded warm inflation theory to encompass a more general noncanonical scalar field model, and further introduced the nonminimal mechanism to analyze the consistency of the new model with observational data, as well as to delve into its unique physical features.

In the multicomponent warm inflation scenario with NMDC, the total action that incorporates all components of the universe is represented as follows:
\begin{equation}
\begin{aligned}
S =& \int d^4x \, \sqrt{-g} \left[ \frac{1}{2} M_p^2 R + \mathcal{L}(X, \phi) \right. \\
&+ \left. \frac{G_{\mu\nu}}{2 M^2} \partial^\mu \phi \partial^\nu \phi + \mathcal{L}_{\rm int} + \mathcal{L}_R \right],
\end{aligned}\label{eq:total}
\end{equation}
where $G_{\mu\nu}$ represents the Einstein tensor, and $M$ is the coupling constant with mass dimension.

The total Lagrangian density is given by
$\mathcal{L}_{\rm total} = \mathcal{L}_g + \mathcal{L}(X, \phi) + \mathcal{L}_{\rm NMDC} + \mathcal{L}_R + \mathcal{L}_{\rm int},$
where $\mathcal{L}_g = \frac{1}{2} M_p^2 R$ denotes the Lagrangian density of the gravitational sector, and $\frac{1}{2M^2} G_{\mu\nu} \partial^\mu \phi \partial^\nu \phi$ signifies the Lagrangian density of the nonminimal derivative coupling term between the inflaton field and gravity.
The Lagrangian density for the noncanonical inflaton field is
$\mathcal{L}_{\rm noncan} = \mathcal{L}(X, \phi),$
which is a general function of the inflaton field $\phi$ and the kinetic term $X$. To maintain the field's canonical normalization, the Lagrangian density must revert to its canonical form (i.e., $\mathcal{L} = X - V$) in the very small $X$ limit.

A crucial parameter that characterizes the propagation speed of scalar perturbations in noncanonical scalar fields is the sound speed, defined as
$c_s^2 = \frac{p_X(\phi,X)}{\rho_X(\phi,X)} = \left(1 + 2X \frac{\mathcal{L}_{XX}}{\mathcal{L}_X}\right)^{-1}$, where a subscript $X$ indicates the first-order derivative with respect to $X$.
The interaction term $\mathcal{L}_{\rm int}$ in Eq.~\eqref{eq:total} relies on the zeroth-order terms of the inflaton and other fields (i.e., the fields themselves), rather than their derivatives. Additionally, the contribution of $\mathcal{L}_{\rm int}$ can be divided into the thermal damping term $\Gamma \dot{\phi}$ and the thermal corrections to the effective potential $V_{\rm eff}$.

Then, the equation of motion for the inflaton field is given by:
\begin{equation}
\left(\mathcal{L}_X c_s^{-2} + 3F \right) \ddot{\phi} + 3H \left(\mathcal{L}_X + 3F + \frac{2 \dot{F}}{H} \right) \dot{\phi} + \Gamma \dot{\phi} + V_{\rm eff,\phi} = 0,
\label{eq:NMDC_inflaton_motion}
\end{equation}
where $F = \frac{H^2}{M^2}$ serves as a dimensionless parameter characterizing the strength of the NMDC. The Einstein GR limit is recovered for $F\ll 1$, whereas $F\gg 1$ represents the high gravitational friction regime.

By integrating the Hamiltonian constraint with the contribution from radiation, the Friedmann equation for the NMDC inflation model can be expressed as:
\begin{equation}
3H^2 = \frac{1}{M_p^2} \left( 2 X \mathcal{L}_X - \mathcal{L} + 9 F X + T s \right).
\label{eq:NMDC_friedmann}
\end{equation}

From the total energy conservation equation of the FRW universe $\dot{\rho} + 3 H (\rho+p) = 0$, and incorporating the field equation of motion Eq.~\eqref{eq:NMDC_inflaton_motion}, the entropy production equation is derived as:
\begin{equation}
T \dot{s} + 3 H T s = \Gamma \dot{\phi}^2.
\label{eq:NMDC_entropy_production}
\end{equation}
By disregarding the highest-order terms in Eqs.~\eqref{eq:NMDC_inflaton_motion} and \eqref{eq:NMDC_entropy_production}, the slow-roll approximation equations are obtained:
\begin{equation}
3 H (\mathcal{L}_X + 3 F) \dot{\phi} + \Gamma \dot{\phi} + V_\phi = 0,
\label{eq:NMDC_slowroll_phi}
\end{equation}
\begin{equation}
3 H T s = \Gamma \dot{\phi}^2.
\label{eq:NMDC_slowroll_entropy}
\end{equation}

The number of e-folds for the noncanonical warm inflation model with nonminimal derivative coupling is represented as:
\begin{equation}
N = \int H \, dt = \int \frac{H}{\dot{\phi}} \, d\phi \simeq - \frac{1}{M_p^2} \int_{\phi_*}^{\phi_{\rm end}} \frac{V (\mathcal{L}_X + 3 F + r)}{V_\phi} \, d\phi.
\label{eq:NMDC_efolds}
\end{equation}

The slow-roll parameters for this model align with those in Eqs.~\eqref{slow-roll_parameters} and \eqref{bc}. Alongside the stability analysis, the sufficient conditions for the validity of the slow-roll approximation are presented as \cite{Zhang2024}:
\begin{equation}
    \begin{aligned}\epsilon&<\mathcal{L}_{X}+3F+r, \quad \beta<\frac{(\mathcal{L}_{X}c_s^{-2}+3F)(\mathcal{L}_{X}+3F+r)}{r},\\ \eta&<\mathcal{L}_{X}c_s^{-2}+3F,\quad
    b<\frac{\mathcal{L}_{X}c_s^{-2}+3F}{\mathcal{L}_{X}+3F+r}.\end{aligned}
\end{equation}
Based on the stability analysis, the constraint $|c| < 4$ is obtained to guarantee that the temperature dependence of the dissipation coefficient does not disrupt the slow-roll evolution of warm inflation.

\vspace{2ex}
\noindent\textbf{Cosmological perturbations}
\vspace{1ex}

Utilizing the field fluctuation relation for warm inflation $\delta \phi^2 = \frac{k_F T}{2 \pi^2}$, where the freeze-out momentum is defined by $k_F = H\sqrt{3(1+Q)}$,
where $Q \equiv r/(\mathcal{L}_X+3F)$ denotes the effective dissipation strength adapted to our NMDC warm DBI inflationary model \cite{Zhang2024}. In the context of spatially flat gauge, with $\mathcal{R} = \frac{H}{\dot{\phi}} \delta \phi_k$, we can obtain the scalar power spectrum of the NMDC noncanonical warm inflation model by integrating the field fluctuation amplitude and the slow-roll equations.
\begin{equation}
\begin{aligned}
P_R = \frac{9 H^5 T \left( \mathcal{L}_X + 3 F + r \right)^{5/2}}{2 \pi^2 V_\phi^2} \sqrt{\frac{3}{\mathcal{L}_X + 3 F}}.
\end{aligned}
\label{eq:scalar_power_spectrum}
\end{equation}
According to the observational constraints of the CMB, the power spectrum is normalized as $P_R \approx 10^{-9}$ on large scales.
The spectral index $n_s$ of the power spectrum is defined as:

\begin{align}
n_s - 1 &= \frac{d \ln P_R}{d \ln k} \simeq \frac{\dot{P}_R}{H P_R} \notag \\
&= \alpha_1 \frac{\epsilon}{\mathcal{L}_X + 3 F + r}
+ \alpha_2 \frac{\eta}{\mathcal{L}_X c_s^{-2} + 3 F}  \label{eq:scalar_spectral_index} \\
& + \alpha_3 \frac{r \beta}{(\mathcal{L}_X c_s^{-2} + 3 F)(\mathcal{L}_X + 3 F + r)}
+ \alpha_4 \frac{\mathcal{L}_X + 3 F + r}{\mathcal{L}_X c_s^{-2} + 3 F} b. \notag
\end{align}
The parameters $\alpha_1$, $\alpha_2$, $\alpha_3$, and $\alpha_4$ are all of order unity. Consequently, in the slow-roll regime, $(n_s - 1)$ is a first-order small quantity with a magnitude of
$\mathcal{O}\left( \frac{\epsilon}{\mathcal{L}_X + 3 F + r} \right) \ll 1.$
Thus, a nearly scale-invariant power spectrum is obtained, which aligns qualitatively with cosmological observations.

Similar to standard inflation, tensor perturbations are independent of the thermal background and generated solely by quantum fluctuations \cite{Taylor2000}. The nonminimal derivative coupling $G^{\mu\nu}\partial_\mu\phi\partial_\nu\phi$ has only a minor impact on the gravitational-wave power spectrum $P_T\simeq \frac{2}{M_p^2}\left(\frac{H}{2\pi}\right)^2$ \cite{Karydas2102.08450,DalianisJCAP2020}. At leading order, the spectral index of tensor perturbations can be expressed as
$n_T = \frac{2 \epsilon}{\mathcal{L}_X + 3 F + r},$
and the tensor-to-scalar ratio $R$ is represented as:
\begin{equation}
R = \frac{P_T}{P_R} = \frac{H}{T} \frac{2 \epsilon (\mathcal{L}_X + 3 F)^{1/2}}{\sqrt{3} (\mathcal{L}_X + 3 F + r)^{5/2}}.
\label{eq:tensor_to_scalar_ratio}
\end{equation}
In the NMDC noncanonical warm inflation scenario, $R$ is greatly suppressed due to the joint influence of nonminimal coupling and thermal dissipation, thereby complying with the current observational upper bound $R < 0.063$ \cite{PLANCK1}.

\section{\label{sec:level3}Theoretical analysis of warm DBI inflation with NMDC}
The warm DBI inflation model was initially introduced and systematically investigated in Ref. \cite{Cai2011}. Stemming from the brane inflation scenario \cite{Dvali1999,Burgess2001,Kachru2003} in warped compactification \cite{Greene2000,Giddings2002,Kachru68}, the DBI inflation model \cite{Alishahiha2004,Loverde2008,Chen2005} allows for more easily satisfied slow-roll conditions in its warm realization compared to canonical cold inflation \cite{Zhang2014}.

This model is categorized into two types \cite{Loverde2008,Chen2005}: the ultraviolet (UV) model, where the inflaton transitions from the UV to the infrared (IR) regime of the potential, and the IR model, where the inflaton transitions from the IR to the UV regime.
The UV model generally yields weaker non-Gaussian signatures compared to the IR case, yet it is less favored by current observations \cite{Planck2014}. In light of these considerations, we consider a straightforward and widely studied power-law potential of the form:
\begin{equation}
V(\phi)=V_0\left(\frac{\phi}{\mu}\right)^n,
\end{equation}
where $V_0>0$ indicates the potential amplitude of dimension $[V_0]=[m]^4$, $n>0$ represents the power-law index, and $\mu$ is a constant scale parameter with mass dimension. In this study, we investigate whether the combined effects of DBI dynamics, NMDC-induced gravitational friction, and thermal dissipation can align this model with current observational constraints within the UV DBI framework.

We primarily focus on two typical values $n=2$ and $n=4$, corresponding to the quadratic potential $V(\phi)=\frac12 m^2\phi^2$ and quartic potential $V(\phi)=\frac14\lambda\phi^4$, respectively. Here, $m$ denotes the inflaton mass, and $\lambda$ is the self-coupling constant.
The quadratic and quartic power-law potentials have been extensively studied and ruled out by cosmological observations within the traditional cold inflation framework \cite{BereraIanRamos2009,PLANCK1,Achucarro2022}.
The incorporation of nonminimal coupling gravitational friction or thermal effects that can generate the dissipation term $\Gamma\dot{\phi}$ allows the $\lambda\phi^4$ potential model, typically excluded in standard minimal-coupling inflation, to remain consistent with the existing cosmological observations \cite{YangNan2015,BeingWarm2014}.

The Lagrangian density for a general inflaton field can be divided into two forms: the series-form Lagrangian and the closed-form Lagrangian \cite{Franche2010}. Under the specific gauge condition \(\mathcal{L}_X=c_s^{-1}\), the latter simplifies to either canonical inflation or DBI inflation \cite{Bean2008,Zhang2015}. The Lagrangian density of the DBI inflaton field is primarily dominated by the warp factor of the AdS-like throat, with its core form expressed as:
\begin{equation}
\mathcal{L}_{DBI} = f^{-1}\left[1 - \sqrt{1 - 2fX}\right] - V(\phi),
\label{eq:DBI_core_Lagrangian}
\end{equation}
where \(f = \Lambda^{-4}\) represents the constant warp factor with \(\Lambda\) the scale parameter \cite{Franche2010}, and this setup reduces the dependence of \(f\) on \(\phi\) \cite{Zhang2014,Franche2010}.

In the thermal dissipation mechanism of warm DBI inflation, we simplify the dissipation coefficient by assuming it to be constant \cite{Berera1995,Yeasmin2023,Alhallak2023,Shiravand2024}, i.e., $\Gamma=\Gamma_0$, leading to $\beta=0$ and $c=0$ in Eqs.~\eqref{slow-roll_parameters} and \eqref{bc}. The slow-roll parameters are as follows:
\begin{equation}
\epsilon=\frac{M_p^2n^2}{2\phi^2}, \quad \eta=M_p^2\frac{n(n-1)}{\phi^2}.
\label{N-B_slow-roll_parameters}
\end{equation}

For the noncanonical inflaton, the first derivative of the Lagrangian density concerning the kinetic term \(\mathcal{L}_X\) satisfies the following relation with the sound speed \(c_s\) of scalar perturbations:
\begin{equation}
\mathcal{L}_X = c_s^{-1} = \left(1-2fX\right)^{-1/2}.
\end{equation}
From the linear stability analysis presented in our earlier work \cite{Zhang2024}, the slow-roll conditions for the present model are expressed as follows:
\begin{equation}
\epsilon\ll c_s^{-1}+3F+r, \quad \eta\ll c_s^{-3}+3F, \quad b\ll 1.
\label{eq:slowroll_conditions_DBI}
\end{equation}
The above equations indicate that the slow-roll conditions derived from our NMDC warm DBI inflationary model are significantly more relaxed than those for both canonical and noncanonical warm inflation within the GR frame, not to mention the standard inflation. This framework allows for a natural realization of slow-roll evolution, independent of a specific inflaton potential. Due to this, the selection of the potential becomes less restrictive, making it possible to incorporate various new models into cosmological inflation.

With respect to the DBI noncanonical relation $\mathcal{L}_X=c_s^{-1}$ in this model, the second-order Langevin equation for inflaton perturbations within our NMDC warm DBI model can be obtained using the methodology established in our previous work \cite{Zhang2024}. By performing a Fourier transform on this second-order Langevin equation, the evolution equation for the fluctuations is derived as:
\begin{equation}
\begin{aligned}
\left(c_s^{-3} + 3F\right)\delta\ddot{\phi}_\mathbf{k} + 3H\left(c_s^{-1} + 3F + r\right)\delta\dot{\phi}_\mathbf{k} \\
+ \left( c_s^{-1}\frac{k^2}{a^2} + 3F\frac{k^2}{a^2} + V_{\phi\phi} \right)\delta\phi_\mathbf{k} = \xi_\mathbf{k},
\end{aligned}
\label{eq:Langevin_Fourier_NMDC_DBI}
\end{equation}
where $\langle\xi(\mathbf{k},t)\xi(-\mathbf{k}',t')\rangle
=2\Gamma T a^{-3}(2\pi)^3\delta^3(\mathbf{k}-\mathbf{k}')\delta(t-t')$ signifies the thermal noise satisfying the standard fluctuation-dissipation theorem \cite{Lisa2004,Gleiser1994}. In the overdamped slow-roll regime, the inertia term can be disregarded, and the equation is simplified to the first-order form:
\begin{equation}
3H\left(c_s^{-1} + 3F + r\right)\delta\dot{\phi}_\mathbf{k} + \left( c_s^{-1}\frac{k^2}{a^2} + 3F\frac{k^2}{a^2} + V_{\phi\phi} \right)\delta\phi_\mathbf{k} = \xi_\mathbf{k}.
\end{equation}

The field value at the end of inflation is derived from the breakdown of the slow-roll condition \(\epsilon=c_s^{-1}+3F+r\), represented as:
\begin{equation}
\phi_{end}^2=\frac{M_p^2n^2}{2\left(c_s^{-1}+3F+r\right)}.
\label{eq:phi_end_NMDC_DBI}
\end{equation}
Substituting the above equation into the integral expression for the number of e-folds, we can rearrange it to derive the inflaton field value at Hubble horizon crossing:
\begin{equation}
\phi_*^2=\frac{2nNM_p^2}{c_s^{-1}+3F+r}+\frac{M_p^2n^2}{2\left(c_s^{-1}+3F+r\right)}.
\label{eq:phi_star_NMDC_DBI}
\end{equation}

The freeze-out physical momentum $k_F = H\sqrt{3(1+Q)}$, represents the critical scale below which inflaton modes cease to be influenced by thermal noise over one Hubble time as the universe expands. Modes with physical wavenumbers greater than $k_F$ can achieve full thermalization, whereas those with smaller ones exit the process of thermal relaxation. For the NMDC warm DBI inflation model in this study, the scalar power spectrum of the model is derived as:
\begin{equation}
\begin{aligned}
P_R&=\left(\frac{H}{\dot{\phi}}\right)^2\delta\phi^2=\frac{H^3T}{2\pi^2u^2}\sqrt{\frac{3\left(\mathcal{L}_X+3F+r\right)}{\mathcal{L}_X+3F}}\\
&=\frac{9H^5T{\mu}^{2n}\left(c_s^{-1}+3F+r\right)^\frac{5}{2}}{2\pi^2{n^2V}_0^2\phi^{2n-2}}\sqrt{\frac{3}{c_s^{-1}+3F}}.
\end{aligned}
\label{eq:scalar_power_spectrum_NMDC_DBI}
\end{equation}

With $\beta=0$ and $b\to0$, the $\alpha_3$ and $\alpha_4$ contributions are discarded, simplifying the scalar spectral index $n_s$ to:
\begin{equation}
\begin{aligned}
n_s-1&=\frac{\mathrm{d}\ln P_R}{\mathrm{d}\ln k}\simeq\frac{\dot{P}_R}{H P_R}\\
&=\frac{\alpha_1 \epsilon}{\mathcal{L}_X+3F+r}+\frac{\alpha_2 \eta}{\mathcal{L}_X c_s^{-2}+3F}\\
&=\frac{\alpha_1 M_p^2n^2}{2\phi_*^2\left(c_s^{-1}+3F+r\right)}+\frac{\alpha_2 M_p^2n(n-1)}{\phi_*^2\left(c_s^{-3}+3F\right)}.
\end{aligned}
\label{eq:spectral_index_NMDC_DBI}
\end{equation}

Tensor perturbations do not interact with the thermal background and are exclusively produced by quantum fluctuations, as is the case in standard inflation \cite{Taylor2000}. By integrating the tensor power spectrum$P_T \simeq \frac{2}{M_p^2}\left(\frac{H}{2\pi}\right)^2$
with Eq.~\eqref{eq:scalar_power_spectrum_NMDC_DBI}, we derive the tensor-to-scalar ratio $R$.
\begin{equation}
R=\frac{P_T}{P_R}=\frac{H}{T}\frac{M_p^2 n^2 \left(c_s^{-1}+3F\right)^\frac{1}{2}}{\sqrt{3}\ \phi_\ast^2 \left(c_s^{-1}+3F+r\right)^\frac{5}{2}}.
\label{eq:tensor_to_scalar_ratio_NMDC_DBI}
\end{equation}
Utilizing the Stefan--Boltzmann relation $\rho_r = \pi^2 g_* T^4 /30$, where $g_*$ represents the effective number of relativistic degrees of freedom and is generally of order $\mathcal{O}(10^2)$, along with the slow-roll equations and the scalar power spectrum from Eq.~\eqref{eq:scalar_power_spectrum_NMDC_DBI}, we obtain the $T/H$ ratio for the NMDC warm DBI inflation model:
\begin{equation}
\frac{T}{H} = \left( \frac{45r}{4\pi^4 g_* P_R} \right)^{\frac{1}{3}} \left[ \frac{3\left(c_s^{-1} + 3F + r\right)}{c_s^{-1} + 3F} \right]^{\frac{1}{6}}.
\end{equation}
This ratio is marginally smaller than that in warm inflation models within the GR limit. An increase in thermal dissipation strength can considerably raise $T/H$. Since $g_*$ is on the order of $\mathcal{O}(10^2)$ during inflation and the scalar power spectrum $P_R$ is of order $\mathcal{O}(10^{-9})$, the warm inflation condition $T>H$ is easily met only when $r>10^{-7}$. Consequently, even very weak thermal dissipation can trigger warm inflation, allowing the model to naturally satisfy the physical prerequisites of warm inflation. Furthermore, thermal fluctuations dominate over quantum fluctuations across the weak dissipation regime, reinforcing the realization of warm inflationary dynamics.

Due to the combined effects of nonminimal derivative coupling and thermal dissipation, the tensor perturbations are notably suppressed. Along with the slow-roll condition $ (\epsilon \ll c_s^{-1}+3F+r) $, the upper bound of the tensor-to-scalar ratio is given by:
\begin{equation}
R < \frac{H}{T}\frac{2}{\sqrt{3}}\frac{1}{\left(c_s^{-1}+3F\right)\left(1+Q\right)^{3/2}}.
\label{eq:tensor_ratio_upper_bound}
\end{equation}
This upper bound indicates that tensor perturbations are significantly diminished in the presence of both nonminimal derivative coupling and thermal dissipation, especially in the strong dissipative regime ($Q \gg 1$), thereby naturally resulting in a small tensor-to-scalar ratio that aligns with observational bounds.

Another important feature of the NMDC warm DBI inflationary model is its effective mechanism for mitigating the $\eta$ problem while maintaining a controlled inflaton field variation. In standard cold inflation, the slow-roll condition necessitates that $\eta \ll 1$, which imposes strict constraints on the curvature of the inflaton potential, making it difficult to realize many simple potentials consistently, such as monomial potentials. Specifically, small-field models, despite involving sub-Planckian field excursions, generally suffer from $|\eta|\sim \mathcal{O}(1)$ due to inadequate flattening of the potential, leading to a violation of the slow-roll condition. Conversely, large-field models can satisfy $|\eta|\ll 1$, but only by requiring super-Planckian field excursions, $\Delta\phi \gg M_p$, which raises questions regarding the validity of the effective field theory framework. In the NMDC warm DBI inflationary scenario, however, the slow-roll condition associated with $\eta$ is adjusted to
$\eta \ll c_s^{-3} + 3F$. According to Eq.~\eqref{eq:phi_star_NMDC_DBI}, the value of $\eta$ at Hubble horizon crossing can be expressed as
\begin{equation}
\eta_*=
\frac{2(n-1)\left(c_s^{-1}+3F+r\right)}
{4N+n}.
\label{eq:eta_star_NMDC_DBI}
\end{equation}
Hence, the effective quantity controlling the validity of the slow-roll approximation is
\begin{equation}
\frac{\eta_*}{c_s^{-3}+3F}
=
\frac{2(n-1)\left(c_s^{-1}+3F+r\right)}
{(4N+n)\left(c_s^{-3}+3F\right)}
\ll 1 .
\label{eq:eta_effective_NMDC_DBI}
\end{equation}
Eq.~\eqref{eq:eta_effective_NMDC_DBI} elucidates the manner in which the $\eta$ problem is mitigated in the present model. In standard cold inflation, the condition $\eta\ll1$ necessitates a sufficiently flat inflaton potential, which severely constrains the allowed potential forms. Conversely, in the NMDC warm DBI scenario, the relevant condition is not $\eta_*\ll1$, but
$\frac{\eta_*}{c_s^{-3}+3F}\ll1$ .
Therefore, the $\eta$ problem is not solved by reducing the potential curvature itself, but by relaxing the effective slow-roll constraint imposed on $\eta$. For $0<c_s<1$, the DBI effect amplifies the factor $c_s^{-3}$ in the slow-roll condition, thus relaxing the constraint on $\eta$, while the NMDC term $3F$ independently increases the gravitational friction, further broadening the viable slow-roll regime. As a result, potentials that might be too steep or too curved in conventional cold inflation can remain viable within this framework.

Another significant theoretical implication of the enhanced damping mechanism is the reduction of the inflaton field excursion.
During the horizon exit of cosmological scales related to multipoles $1 \leq l \leq 100$, the corresponding e-folding interval is roughly $\Delta N \simeq 4.6$.
Using the slow-roll equation of motion, the field variation during this interval can be approximated as
\begin{equation}
\begin{aligned}
\frac{\Delta\phi}{M_p}
&= \frac{\dot{\phi}\Delta N}{M_p H}
\lesssim
\frac{4.6\sqrt{2}}
{\left(c_s^{-1} + 3F + r\right)^{1/2}}  \\
&\sim
\left(c_s^{-1} + 3F + r\right)^{-1/2}.
\end{aligned}
\label{eq:field_excursion_NMDC_DBI}
\end{equation}
This expression indicates that the field excursion is governed by the same effective damping factor that affects the slow-roll dynamics.
The DBI noncanonical kinetic contribution, the NMDC-induced gravitational friction, and the thermal dissipative effect all serve to increase the denominator in Eq.~\eqref{eq:field_excursion_NMDC_DBI}, which consequently diminishes the required inflaton displacement during observable inflation.

As a result, the conventional Lyth bound obtained in standard cold inflation is significantly altered within the present NMDC warm DBI inflationary framework.
In cold large-field inflation, an observationally relevant tensor-to-scalar ratio is typically associated with a Planckian or super-Planckian field excursion.
In contrast, in the present model, the tensor-to-scalar ratio is already notably reduced due to thermal fluctuations and heightened friction, while the inflaton excursion is simultaneously suppressed by the factor
$\left(c_s^{-1}+3F+r\right)^{-1/2}$.
Hence, the model allows for sufficient inflation with a sub-Planckian field variation, $\Delta\phi\ll M_p$, without requiring super-Planckian inflaton motion.
This offers a natural method to circumvent the super-Planckian field excursion difficulty typically found in conventional large-field cold inflation models.

The integration of NMDC and DBI kinetic structure establishes a framework that can effectively tackle two key challenges of conventional warm inflation models. The interaction between NMDC-induced gravitational friction and DBI dynamics not only relaxes the $\eta$ slow-roll constraint but also suppresses the inflaton field excursion to sub-Planckian scales. Consequently, the model realizes a fully self-consistent warm inflation scenario with enhanced theoretical consistency and improved agreement with observational constraints.

\vspace{1ex}
\noindent\textbf{Dynamical regimes}
\vspace{1ex}

The background evolution of the NMDC warm DBI inflation model is primarily influenced by the thermal dissipation strength parameter $r$ and the relative significance of the NMDC coupling strength parameter $F$. Consequently,
different choices of $(r,F)$ lead to various effective damping mechanisms.
For clarity, the parameter space can be qualitatively categorized into the following representative dynamical regimes.

\begin{itemize}
    \item \textbf{NMDC-dominated regime:}
    $r \lesssim 10^{-7}$ and $F=\mathcal{O}(10^2\sim 10^3)$.
    In this regime, thermal dissipation is minimal, while the NMDC-induced gravitational friction serves as the primary source of damping.
    The slow-roll evolution is predominantly maintained by the enhanced gravitational friction, allowing the model to effectively resemble a cold DBI inflation scenario with NMDC.

     \item \textbf{Thermal-dominated regime:}
    $r=\mathcal{O}(10^2\sim 10^3)$ and $F \lesssim 10^{-3}$.
    In this regime, the background dynamics is primarily influenced by thermal dissipation, whereas the NMDC-induced gravitational friction is negligible.
    The model thus simplifies to the warm DBI inflationary scenario in the GR limit.
    Compared to the balanced or strongly overdamped regimes, this case offers a more restricted viable parameter space since the lack of NMDC friction weakens the regulation of the scalar spectral index.

     \item \textbf{Strongly overdamped regime:}
     $r=\mathcal{O}(10^3)$ and $F=\mathcal{O}(10^3)$.
     In this regime, both the thermal dissipation and the NMDC-induced gravitational friction are dominant, resulting in a strongly overdamped evolution of the inflaton field. Consequently, the effective friction term is greatly enhanced, which relaxes the slow-roll conditions and stabilizes the inflationary dynamics.

     This regime typically achieves the most substantial suppression of the tensor-to-scalar ratio, underscoring the cooperative interaction between warm dissipation and NMDC effects in controlling the inflationary evolution.

    \item \textbf{GR-based critical warm inflation regime:}
    $r \lesssim 10^{-7}$ and $F \lesssim 10^{-3}$.
    In this regime, both the NMDC gravitational friction and the thermal dissipative effect are significantly suppressed.
    The model approaches the GR-based DBI inflationary limit.
    When $r$ is of order $\mathcal{O}(10^{-7})$, the thermal bath approaches the threshold necessary to maintain the warm inflation condition, so this region is referred to as the critical warm inflation regime.

    \item \textbf{Balanced regime:}
    $r=\mathcal{O}(10\sim 10^2)$ and $F=\mathcal{O}(10\sim 10^2)$.
    In this regime, thermal dissipation and NMDC gravitational friction are roughly equivalent in magnitude, and neither effect solely dominates the background dynamics.
    The inflaton evolution is jointly influenced by the two damping mechanisms, resulting in an intermediate regime where both effects contribute appreciably to the slow-roll behavior and to the observational predictions.
\end{itemize}

\section{\label{sec:level4} Comparison with observations}
This study uses the \emph{Planck} 2018 CMB data \cite{PLANCK1}, including the temperature-temperature (TT) angular power spectrum, the temperature--E-mode polarization (TE) cross-power spectrum, the E-mode polarization (EE) angular power spectrum, the low-$\ell$ E-mode polarization (lowE) likelihood, and the CMB lensing reconstruction. These data are supplemented by the B-mode polarization measurements from the 2015 BICEP2/Keck Array (BK15) and baryon acoustic oscillation (BAO) data \cite{Beutler2011,Ross2015,Alam2017}. The resulting observational constraints are presented at the 68\% and 95\% confidence levels.

The coefficients \(\alpha_1\) and \(\alpha_2\), which appear in the formulation of \(n_s\), originate from the thermal fluctuation dynamics and the freeze-out process; thus, they are not free parameters but are determined by the background evolution. For a constant dissipation coefficient (\(c=0\)), their leading behavior can be estimated as
\begin{equation}
\alpha_1 \simeq -3 + \mathcal{O}\!\left(\frac{r}{\mathcal{L}_X+3F}\right),\qquad
\alpha_2 \simeq \frac{6}{\Delta} + \mathcal{O}\!\left(\frac{r}{\mathcal{L}_X+3F}\right),
\end{equation}
with \(\Delta = 4\bigl(1+\frac{r}{\mathcal{L}_X c_s^{-2}+3F}\bigr)\). Both coefficients are of order unity.
Their precise values are contingent on the full dynamics; however, their contributions to (\(n_s-1\)) take the form \(\alpha_1 A + \alpha_2 B\), with the prefactors \(A\) and \(B\) being slow-roll suppressed and typically of order \(\mathcal{O}(10^{-4}\sim 10^{-3})\) in the parameter region of interest. As a result, moderate changes in \(\alpha_1,\alpha_2\) result in only subleading changes in the observables, and the predictive power of the model is predominantly influenced by the dynamical parameters \(c_s\), \(r\), and \(F\).

While $\alpha_1$ and $\alpha_2$ do not explicitly rely on the potential index $n$, they gain an implicit dependence through the background field value at horizon crossing and the corresponding slow-roll trajectory. Consequently, varying potentials may be related to slightly different effective values. Within the parameter space permitted by the stability analysis, the slow-roll conditions, the typical ranges of the sound speed $c_s$, the NMDC coupling
strength parameter $F$, and the thermal dissipation strength parameter $r$, we observe typical values $(\alpha_1,\alpha_2)=(-3.8,-0.2)$ for the quadratic potential ($n=2$) and $(-1.9,-0.07)$ for the quartic  potential ($n=4$). These values are not introduced as additional tunings; rather, they simply reflect the typical regions chosen by the combined dynamical consistency and theoretical constraints. We have explicitly confirmed that moderate fluctuations around these values do not change the qualitative conclusions, reinforcing the robustness of our results.

With this foundation, we now proceed to examine the observational predictions of the model, emphasizing the dependence of $n_s$ and $R$ on the relevant parameters.

\vspace{1ex}
\noindent\textbf{Relation $\boldsymbol{n_s\sim c_s}$}
\vspace{1ex}

\begin{figure*}[htbp]
    \centering

    \includegraphics[width=1.0\textwidth]{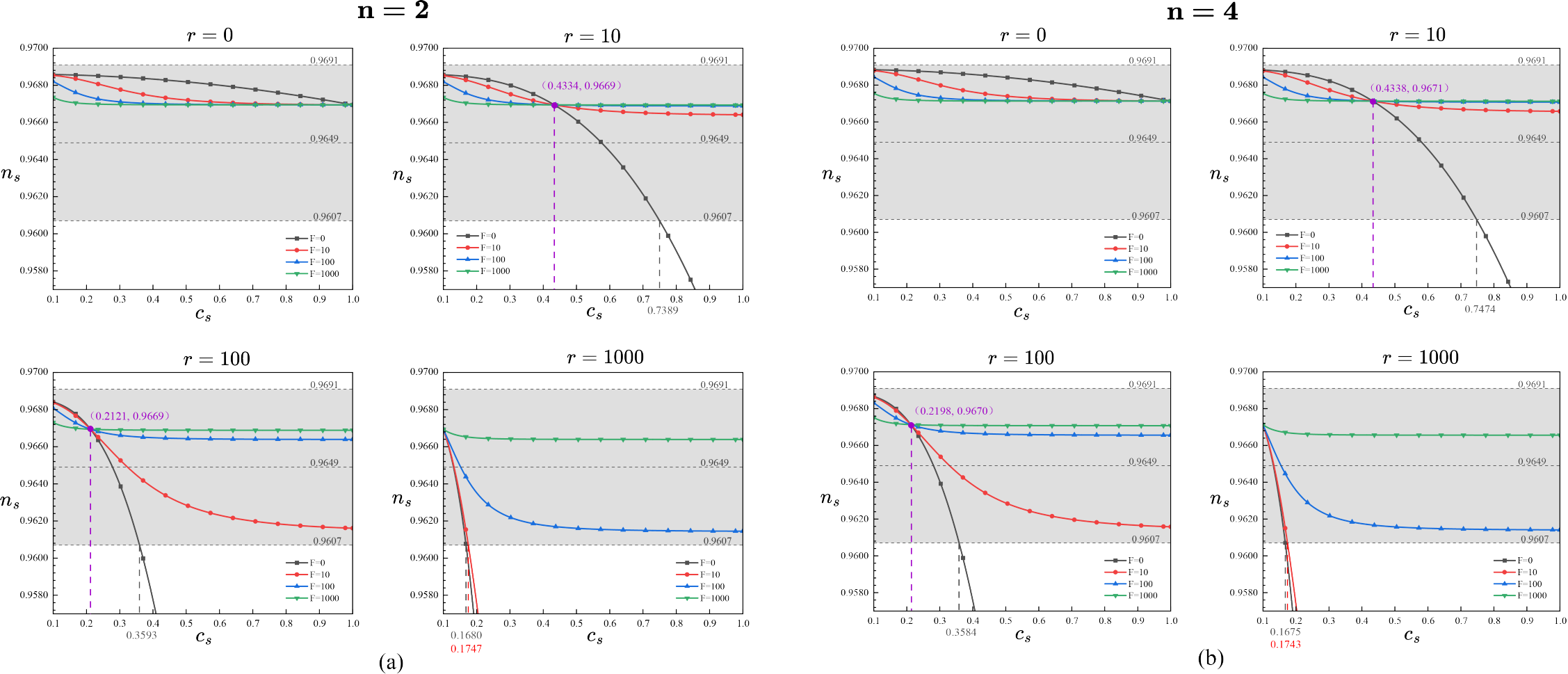}

    \caption{The spectral index $n_s$ of the power spectrum plotted against the sound speed $c_s$ for various values of $F$ and $r$, with $N=60$. (a) $n=2$, $\alpha_1=-3.8$, $\alpha_2=-0.2$, with the \textit{Planck} 2018 68\% CL constraint on $n_s$ ($n_s=0.9649\pm0.0044$) represented by the gray shaded region. (b) $n=4$, $\alpha_1=-1.9$, $\alpha_2=-0.07$. In both panels, the lines correspond to $F=0,10,100,1000$ (black, red, blue, green), and subplots display $r=0,10,100,1000$ from top-left to bottom-right.}
    \label{fig:ns_cs_ns}
\end{figure*}

The scalar spectral index $n_s$ is assessed at Hubble horizon crossing, with the corresponding inflaton field $\phi_*$ given by Eq.~\eqref{eq:phi_star_NMDC_DBI}. By substituting $\phi_*$ into Eq.~\eqref{eq:spectral_index_NMDC_DBI} and setting $M_\mathrm{p} = 2.435 \times 10^{18}~\mathrm{GeV}$, we derive the ($n_s\sim c_s$) trajectories for various combinations of $F$ and $r$, as illustrated in Fig.~\ref{fig:ns_cs_ns}. These curves distinctly exhibit the variation of $n_s$ with $c_s$, along with the joint effects of the NMDC coupling strength parameter $F$ and the thermal dissipation strength parameter $r$.

 Both panels adopt $N=60$ e-folds. The thermal dissipation strength parameter $r$ takes typical values of 0, 10, 100, and 1000 in each subplot. The differently colored curves indicate various NMDC coupling strengths from weak to strong: $F=0$ (black), $F=10$ (red), $F=100$ (blue), and $F=1000$ (green). The gray shaded region signifies the 68\% CL constraint on $n_s$ from \emph{Planck} 2018, i.e., $n_s=0.9649 \pm0.0044$ \cite{PLANCK1}.

Fig.~\ref{fig:ns_cs_ns} illustrates that $n_s$ consistently decreases as sound speed increases, independent of the values of the NMDC coupling strength parameter $F$ and thermal dissipation strength parameter $r$. Observationally viable inflationary models favor a red-tilted scalar spectrum with $n_s < 1$, with smaller $n_s$ indicating a stronger red tilt. The findings reveal that increasing $c_s$ systematically amplifies the red tilt, and this behavior is largely unaffected by both gravitational friction and thermal dissipation.

For $r=0$ (indicating no thermal dissipation), all $n_s$ curves corresponding to different $F$ exhibit a gradual decline with $c_s$ and remain within the \emph{Planck} 68\% CL bounds without any sudden drops. This behavior, in contrast to the cases with higher dissipation strengths ($r=10,\,100,\,1000$) presented in the following panels, clearly indicates that thermal dissipation plays a crucial role in the rapid decrease of $n_s$, which can drive it beyond the observationally favored region. At a particular sound speed, a stronger NMDC coupling strength further drives $n_s$ toward the central value.

For $r=10$ and $r=100$, $n_s$ exhibits a distinct dependence on the speed of sound $c_s$ as the NMDC coupling strength parameter $F$ varies. When the speed of sound $c_s$ is below the critical threshold ($c_s \approx 0.4334, n_s \approx 0.9669$ for $n=2, r=10$; $c_s \approx 0.2121, n_s \approx 0.9669$ for $n=2, r=100$; $c_s \approx 0.4338, n_s \approx 0.9671$ for $n=4, r=10$; $c_s \approx 0.2198, n_s \approx 0.9670$ for $n=4, r=100$), $n_s$ decreases gradually as $F$ increases. Conversely, when $c_s$ surpasses the critical value, $n_s$ gradually rises with increasing $F$. In the strong dissipation case of $r=1000$, $n_s$ consistently increases with $F$ at a fixed speed of sound across the entire range of $c_s$.

Moreover, for $r=10,100,1000$, indicating the presence of thermal dissipation, there exists a critical sound speed $c_s^*$. Once $c_s$ exceeds this critical value, the scalar spectral index $n_s$ moves beyond the \emph{Planck} 68\% CL region. The critical sound speed $c_s^*$ declines as the dissipation strength $r$ rises. For $n=2$, we find $c_s^* \approx 0.7389$ for $r=10$; $c_s^*\approx 0.3593$ for $r=100$; $c_s^* \approx 0.1680$ for $r=1000, F=0$ and $c_s^* \approx 0.1747$ for $r=1000, F=10$. A similar set of critical values is acquired for the quartic potential $(n=4)$, but with slightly different numerical values. This suggests that the potential index $n$ has only a minimal impact on the critical sound speed, while the existence and location of $c_s^*$ are primarily influenced by $r$ and $F$. For $F=0$ and $F=10$, increasing both $r$ and $c_s$ progressively drives the scalar spectral index away from the observationally allowed region. The underlying reason is that, within this parameter range, the slow-roll parameter $\eta$ fails to meet the condition given in Eq.~\eqref{eq:slowroll_conditions_DBI}. As a result, the slow-roll approximation ceases to hold.

As the thermal dissipation strength parameter $r$ rises, the $n_s$ associated with low NMDC coupling strength $F$ declines sharply, potentially falling below the \emph{Planck} 68\% CL observational bounds. Nevertheless, further increasing the NMDC coupling strength parameter $F$ can bring $n_s$ back into the observationally allowed range. This suggests a notable competitive interaction between $F$ and $r$. While thermal dissipation tends to lower $n_s$, resulting in a redder spectrum, $F$ mitigates this effect, raising $n_s$ back into the allowed range. This competition effectively regulates the model and plays a crucial role in determining compliance with observational constraints, significantly expanding the viable parameter space.

For large $F$ values (e.g., $F=1000$), $n_s$ approaches a state that is nearly unaffected by $c_s$ and stabilizes on a plateau slightly above the \emph{Planck} central value $n_s \approx 0.9649$. This illustrates that extremely strong $F$ can diminish the dependence of $n_s$ on the sound speed.

\vspace{2ex}

Overall, the analysis of $(n_s\sim c_s)$ reveals a clear regulatory mechanism in the NMDC warm DBI inflation model. The sound speed affects the red tilt of the scalar spectrum, while thermal dissipation tends to decrease $n_s$ and may drive it beyond the observationally favored region. Conversely, the NMDC-induced gravitational friction can counterbalance this dissipative effect and stabilize $n_s$ across a wide range of sound speeds. Therefore, this competitive interaction between $r$ and $F$ is vital for maintaining observational viability and significantly broadens the allowed parameter space of the model.

\vspace{1ex}
\noindent\textbf{Relation $\boldsymbol{n_s\sim R}$}
\vspace{1ex}

\begin{figure*}[htbp]
    \centering

    \includegraphics[width=1.0\textwidth]{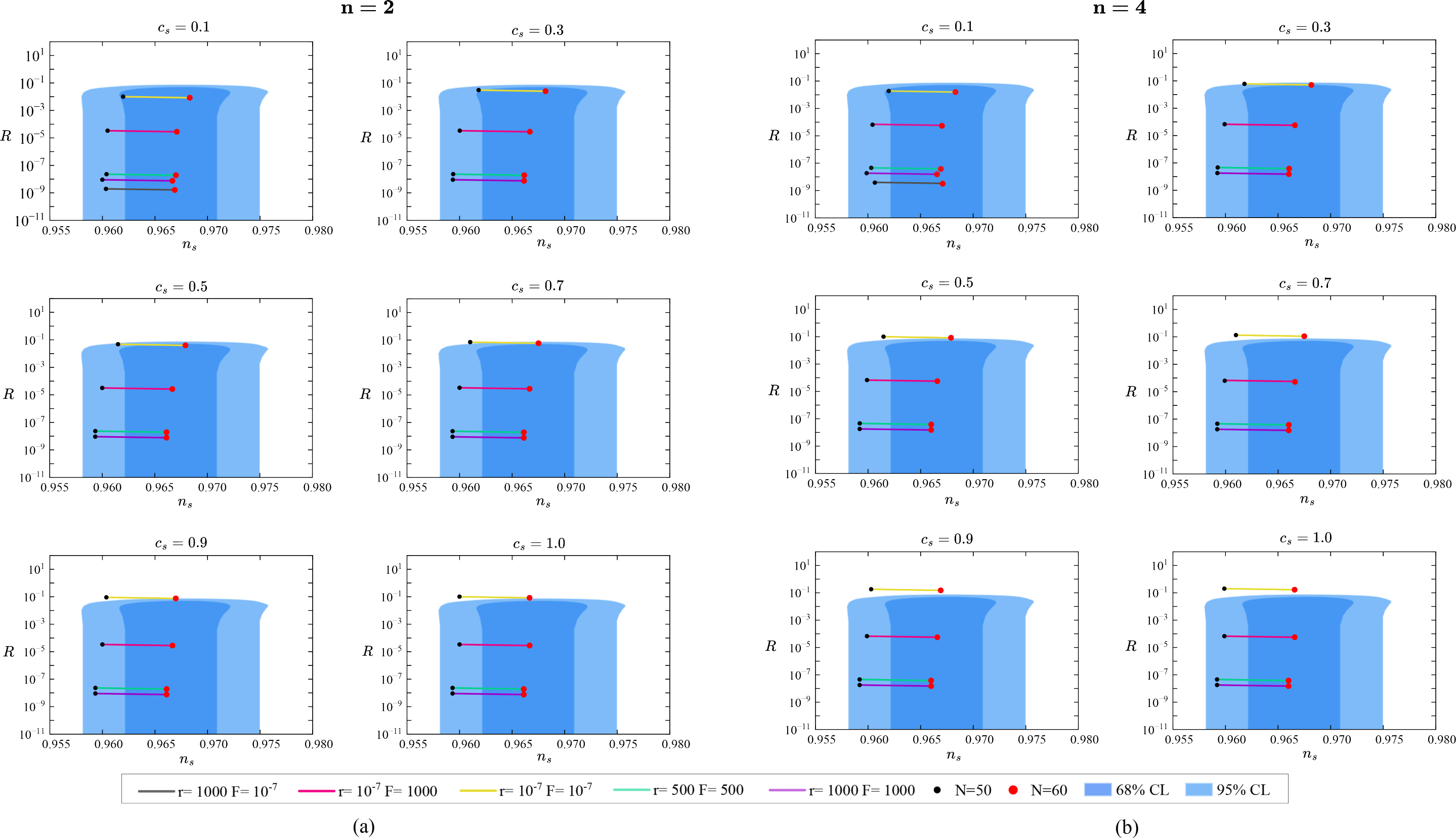}

    \caption{The evolution of the tensor-to-scalar ratio $R$ with the scalar spectral index $n_s$ is depicted for two sets of potential parameters: (a) $n=2$ ($\alpha_1=-3.8$, $\alpha_2=-0.2$) and (b) $n=4$ ($\alpha_1=-1.9$, $\alpha_2=-0.07$). Each panel corresponds to fixed sound speeds $c_s = 0.1, 0.3, 0.5, 0.7, 0.9,$ and $1.0$. The various colored curves illustrate different combinations of the NMDC coupling strength parameter $F$ and the thermal dissipation strength parameter $r$: gray ($r=1000, F=10^{-7}$), pink ($r=10^{-7}, F=1000$), yellow ($r=10^{-7}, F=10^{-7}$), green ($r=500, F=500$), and purple ($r=1000, F=1000$). The black and red markers represent results for e-folding numbers $N=50$ and $N=60$, respectively. The dark- and light-blue shaded regions denote the 68\% and 95\% confidence regions, respectively, from the combined \emph{Planck} 2018+BK15+BAO data.}
    \label{fig:ns_R}
\end{figure*}

Utilizing Eqs.~\eqref{eq:spectral_index_NMDC_DBI} and ~\eqref{eq:tensor_to_scalar_ratio_NMDC_DBI}, we derive the relationship between the scalar spectral index $n_s$ and the tensor-to-scalar ratio $R$, with $g_* = 106.75$ and $P_R = 10^{-7}$. The vertical axis is plotted on a base-10 logarithmic scale to clearly illustrate the wide-range magnitude of $R$. Fig. \ref{fig:ns_R} displays the evolution of $R$ as a function of $n_s$ for two sets of potential parameters that take the same values as before: $n=2$ and $n=4$. The analysis considers sound speeds $c_s=0.1,0.3,0.5,0.7,0.9,$ and $1.0$, alongside various values of the NMDC coupling strength parameter $F$, the thermal dissipation strength parameter $r$, and e-folding numbers $N=50$ and $60$. The dark-blue and light-blue shaded regions denote the 68\% and 95\% confidence regions, respectively, from the combined \emph{Planck} 2018+BK15+BAO constraints in the $(n_s \sim R)$ plane \cite{PLANCK1,Beutler2011,Ross2015,Alam2017}.

As depicted in Fig.~\ref{fig:ns_R}, each colored curve (gray, pink, yellow, green, and purple) signifies a specific parameter selection within the distinct dynamical regimes defined previously, characterized by various combinations of the dissipation strength $r$ and the NMDC coupling $F$. Notably, the parameter sets $(r,F)=(10^{-7},1000)$, $(500,500)$, and $(1000,1000)$ correspond to regimes where NMDC effects and/or thermal dissipation are pronounced, all yielding predictions consistent with the \emph{Planck} 2018+BK15+BAO constraints. The interaction of NMDC-induced gravitational friction and thermal dissipation markedly reduces the tensor-to-scalar ratio $R$, leading to typical values within the range $10^{-8}\lesssim R\lesssim 10^{-5}$. More specifically, for $N=50$, the predicted points fall within the 95\% CL region of \emph{Planck} 2018+BK15+BAO, while for $N=60$, they further enter the 68\% CL region. The remaining yellow curve illustrates the standard cold DBI inflation limit in general relativity, which corresponds to $(r,F)=(10^{-7},10^{-7})$. In this GR limit, as the sound speed $c_s$ increases, the scalar spectral index $n_s$ gradually diverges from the \emph{Planck} 2018+BK15+BAO bounds. This categorization provides a clear framework for comparing the effects of thermal dissipation and NMDC across different dynamical regimes.

In the case dominated by strong thermal effects, with $r=1000$ and $F=10^{-7}$ (represented by the gray curve), the scalar spectral index $n_s$ for both $N=50$ and $N=60$ falls within the 68\% CL region of \emph{Planck} 2018+BK15+BAO at $c_s=0.1$. Nevertheless, for larger sound speeds $c_s=0.3,\,0.5,\,0.7,\,0.9,$ and $1$, the slow-roll parameter $\eta$ no longer meets the condition outlined in Eq.~\eqref{eq:slowroll_conditions_DBI}. Consequently, $n_s$ declines monotonically and moves away from the observationally favored region. This suggests that the purely thermal-dominated scenario offers only a limited viable parameter space, constrained to very low sound speeds. In contrast, the inclusion of NMDC-induced gravitational friction is vital, and the interaction between $F$ and $r$ notably broadens the viable parameter space of the model.

\vspace{2ex}

Based on the above analyses of Figs.~\ref{fig:ns_cs_ns} and \ref{fig:ns_R}, it is observed that the scalar spectral index is primarily influenced by the combined effects of the sound speed, NMDC friction, and thermal dissipation. While thermal dissipation tends to make the scalar spectrum redder, a sufficiently large NMDC friction parameter can counterbalance this effect, ensuring that $n_s$ remains within the observationally favored region. The tensor-to-scalar ratio is significantly suppressed for the representative viable parameter sets, typically falling within the range of $10^{-8}\lesssim R\lesssim 10^{-5}$. Overall, these findings indicate that the cooperative and competitive interaction between DBI kinetics, NMDC friction, and warm dissipation is crucial for achieving observationally viable inflationary predictions.

\section{\label{sec:level5}Conclusions and discussions}
In this paper, we have developed a warm DBI inflationary model with nonminimal derivative coupling to gravity. The model integrates the DBI noncanonical kinetic structure, gravitational friction induced by the NMDC coupling, and thermal dissipation into a cohesive framework. Utilizing this framework, we examined the background dynamics and primordial perturbations, and applied the derived predictions to power-law potentials with $n=2$ and $n=4$.

The key finding of this research is that the interaction between NMDC-induced gravitational friction and thermal dissipation offers an efficient dynamical mechanism for regulating inflationary observables. In the ($n_s \sim c_s$) analysis, the sound speed governs the red tilt of the scalar spectrum, while thermal dissipation tends to lower $n_s$, potentially pushing it beyond the observationally favored region. The NMDC coupling counteracts this effect through enhanced gravitational friction, enabling $n_s$ to remain compatible with the \emph{Planck} 2018 constraints across a broader range of parameters. This complementary interplay between the thermal dissipation strength $r$ and the NMDC coupling strength parameter $F$ forms a crucial dynamical mechanism that expands the viable parameter space of the model. In the $(n_s \sim R)$ plane, the representative parameter sets with NMDC and thermal dissipation align with the observational bounds: the predictions for $N=50$ fall within the 95\% CL region, while those for $N=60$ further enter the 68\% CL region. Meanwhile, the tensor-to-scalar ratio is strongly suppressed, typically reaching $10^{-8}\lesssim R\lesssim 10^{-5}$.

The theoretical importance of the model arises from the fact that the same damping mechanism enhances the internal consistency of warm DBI inflation. The DBI kinetic structure alters the scalar sound speed, the NMDC term increases gravitational friction, and thermal dissipation further dampens the inflaton motion through the radiation bath. Their joint effect relaxes the slow-roll conditions, particularly the effective $\eta$ constraint, thereby addressing the $\eta$ problem. Additionally, the increased damping limits the inflaton field excursion, enabling the model to achieve sufficient inflation with $\Delta\phi\ll M_p$, thus avoiding the super-Planckian field variations typically associated with conventional large-field cold inflation. These characteristics indicate that warm DBI inflation with NMDC can be both theoretically self-consistent and observationally viable.

Several potential extensions of the present work can be considered. A natural next step is to evaluate primordial non-Gaussianity in this NMDC warm DBI framework, as DBI-type models might produce nontrivial higher-order correlation functions. It would be beneficial to move beyond the constant-$\Gamma$ approximation and investigate dissipative coefficients with explicit field or temperature dependence. Addressing these issues could further elucidate the microphysical origin of the dissipative sector and offer additional observational tests of the model.

\section{Acknowledgments}
This work was supported by projects ZR2021MA037 and ZR2022JQ04, supported by Shandong Provincial Natural Science Foundation, and the National Natural Science Foundation of China (Grant No. 12575134 and No. 11605100).

\end{document}